\documentclass[%
 reprint,
 amsmath,amssymb,
 aps,
]{revtex4-2}

\usepackage{graphicx}
\usepackage{dcolumn}
\usepackage{bm}

\begin{document}


\title{Lagrange point stability for a rotating host mass binary}

\author{Martin D. Strong}
\email{mdstrong.astrphys@gmail.com}
 \affiliation{Department of Physics and Astronomy, Louisiana State University,
Baton Rouge, LA 70803\\}
\author{Michael Crescimanno}%
 
\affiliation{Department of Physics and Astronomy, Youngstown State University, Youngstown, OH 44555\\
}


\date{\today}

\begin{abstract}
In this new era of gravitational wave astrophysics, observations 
have indicated the likely existence of black holes with significant spin. In order to
better understand the potential imprint orbital dynamics have on the multi-messenger data, we include rotation of the primary mass to leading order in the analysis of the stability boundary pertaining to the triangular equilibrium points, $L_4$ and $L_5$, in the relativistic, restricted, circular three body problem. For Lagrange point stability these rotation effects are of the same order as the leading order relativistic corrections ignoring rotation and make both $L_4$ and $L_5$ more stable for retrograde orbital motion.

\end{abstract}

\maketitle


\section{\label{sec:1} Introduction}
The recent advent of gravitational wave astronomy and a maturing multi-messenger methodology have contributed some urgency for a more thorough understanding of multi-center relativistic orbital systems.
One important difference relativity introduces into the Newtonian class of problems is that the causal structure, and therefore dynamics, depend on the spin of objects, not just their masses. Incorporating the effects of a gravitating mass's angular momentum into spacetime structure and computing the orbital consequences on a test mass is well understood theoretically \cite{LT,Kerr}. Furthermore, recent gravitational wave events \cite{BHSpinLIGO} have also provided experimental signatures consistent with the coalescence and formation of rapidly spinning compact objects which appear to be near the upper limits predicted by relativity. The planar relativistic two-body problem has also been extensively studied theoretically \cite{Robertson,Contopolous,C&C, DamourBuonanno}, and its numerical applications, including gravitational radiation, play key roles in interpreting experimental gravitational wave data. 

Studying the classical restricted (planar, third mass a test mass) three body problem \cite{lagrange} is valuable for extending one's intuition, particularly for insight into the (leading order) general relativistic context \cite{MaindlDvorak, Krefetz, HW, Sicardy, greeks, YA, IYA, Schnittman, SinghBello}. Our goal here is to elucidate how angular spin momentum, $J$, of the `host' mass, $M$, influences the stability of the equilateral Lagrange points, $L_4$ and $L_5$, in the circular relativistic restricted three body system. Understanding this case is of interest, as declared at the conclusion of the recent paper \cite{Schnittman}, (Pg. 47) ``Coupled with the non-degenerate orbital frequencies of test particles in a Kerr background, the inclusion of spinning BHs would introduce many new degrees of freedom that may affect the stability of $L_4$ and $L_5$."

The following article summarizes an approach and findings regarding this question. 
The main conclusion, limited to a binary system where the primary's (the larger mass's) spin axis is orthogonal to the orbital plane, is that $L_4$ and $L_5$'s stability is degraded by prograde orbital motion, but enhanced in the case of retrograde. Beyond being a new observation of potential relevance pertaining to accretion signals originating from compact binary systems, we suggest that the effect of rotation of the masses on the orbital stability of these admittedly idealized systems (planar, near circular orbits) fits into a larger narrative.

In section II, we review relevant prior work regarding the relativistic restricted three body problem and describe how to extend it to the case of a rotating primary. This leads to a pair of non-linear differential equations which we integrate numerically and, specializing to the linear response theory about the equilibrium points,  organize and summarize their stability criteria in section III. Section IV concludes by providing a brief analytical narrative that places this stability result and others for the restricted 2-body relativistic case in context. 

\section{\label{sec:2} Analytical Description} 
In order to investigate the stability of the Lagrange points in the relativistic, restricted, three-body problem, we begin with the approximate (to leading order, e.g. neglecting gravitational radiation) two-body relativistic system. The existing literature has several formulations of this system \cite{Robertson,Contopolous,C&C, DamourBuonanno, MaindlDvorak, Krefetz, HW, Sicardy, greeks}, although the vast majority do not include the effects of host mass rotation. We work to the leading order of small mass ratio, $m/M<<1$, by modifying the equations of motion (EOM) of Huang and Wu \cite{HW} to include the angular momentum $J$ of $M$. In this limit, and for small $a = J/(Mc)$, the inclusion of host mass rotation changes the binary's orbital period. Furthermore, for a test mass located at $L_4$ and $L_5$, we find additional $a$-dependent terms to be included into the EOM of \cite{HW}. These terms change the location and stability criteria for the Lagrange points. Proceeding earlier literature, the orbital equations are rendered in a rotating (about the center of momentum) Cartesian co-ordinate frame with rate 
\begin{equation}
\Omega_{\pm} = \omega_0+\omega_1/c^2 \pm |a|\omega_0^2
\label{orbitalRate}
\end{equation}
where, hereinafter, the upper and lower sign corresponds to prograde and retrograde orbital motion, respectively.
We also define the dimensionless quantities $\mu_1 = \frac{M}{M+m}$ and $\mu_2 = \frac{m}{M+m}$ and scale $G$ such that the Keplerian rate for the binary is $\omega_0^2  = 1/R^3$ while the leading relativistic correction with $a=0$ is $\omega_1 = (\mu_1\mu_2 - 3)/(2R)$ (see \cite{HW}). 

Our starting point is Eqs.~(13-16) from Huang and Wu \cite{HW} which are consistent with the 1PN equations in \cite{D&P} and \cite{MaindlDvorak}. To orient the reader, these equations do not include rotation (i.e. $a=0$) of the host body, no effects from gravitational radiation or frictional forces, nor any other perturbations. Furthermore, the binary is assumed to occupy a circular orbit. In the rotating Cartesian center-of-momentum co-ordinates, the host mass and secondary are located at $(X_1,0)$ and ($X_2,0)$, respectively, with separation $R=|X_1-X_2|$. For completeness, we reproduce here the equation set (Eqs.~12-16) of \cite{HW} for an infinitesimal third test mass
\begin{equation}
{\ddot X} -2\Omega{\dot Y}-\Omega^2 X = 2\omega_0^2\omega_1/c^2 X-A_{0,3}X+A_{1,3} + P/c^2
\label{EqX}
\end{equation}
\begin{equation}
{\ddot Y} +2\Omega{\dot X}-\Omega^2 Y = -2\omega_0^2\omega_1/c^2 Y -A_{0,3}Y + Q/c^2
\end{equation}
where we have implemented the set of constants $A_{n,m} = \frac{\mu_1X_1^n}{d_1^m} + \frac{\mu_2X_2^n}{d_2^m}$ with $d_{1,2} = \sqrt{(X-X_{1,2})^2+Y^2}$.
The remaining post-Newtonian corrections in $P$ and $Q$ to this order (Eqs.~(15,16) from \cite{HW}) are
\begin{widetext}
\begin{eqnarray}
P=6\omega_0A_{0,1}&&({\dot Y}+\frac{\omega_0}{2}X)+\frac{\mu_1\mu_2}{R}(\frac{X-X_1}{d_1^3}+\frac{X-X_2}{d_2^3})+\omega_0(4{\dot Y} + \frac{7\omega_0}{2}X)(A_{1,3}X-A_{2,3})-\frac{7\omega_0^2}{2}A_{1,1}
\nonumber \\
&&+(4A_{0,1}-(U^2+2A\omega_0+{\cal R}^2\omega_0^2) +4{\dot X}^2-5\omega_0{\dot X}Y+\omega^2Y^2 )(A_{0,3}X-A_{1,3})-3\omega_0A_{0,1}(2{\dot Y}+\omega_0X) \nonumber \\
&&
+A_{0,3}(4{\dot Y}+\omega_0 X)({\dot X}-\omega_0 Y)Y -\frac{3\omega_0^2}{2}(A_{2,3}X-A_{3,3})
+\frac{3\omega_0^2}{2}Y^2(A_{2,5}X-A_{3,5})
\label{prel}
\end{eqnarray} 

\begin{eqnarray}
Q=-A_{1,3}(&&3{\dot X}{\dot Y}+7\omega_0X({\dot X} - \frac{\omega_0}{2}Y)) + \omega_0A_{2,3}(4{\dot X}-\frac{5\omega_0}{2} Y)
+\frac{\mu_1\mu_2}{R} Y(\frac{1}{d_1^3}+\frac{1}{d_2^3}) 
-({\dot Y}-\omega_0X)(\omega_0 Y-{\dot X})(A_{0,3}X-A_{1,3})\nonumber \\ 
&&+ A_{2,5}\frac{3\omega_0^2}{2} Y^3 
+ A_{0,3}\bigl[Y(4A_{0,1}-(U^2+2{\cal A}\omega_0+{\cal R}^2\omega_0^2)
+({\dot Y}+\omega_0 X)^2) + 3({\dot Y}+\omega_0 X)(X{\dot X}+Y{\dot Y}) \bigr]
 \label{qrel}
\end{eqnarray}
\end{widetext}
where, as per \cite{HW}, 
$U^2 = {\dot X}^2+{\dot Y}^2$, ${\cal A} = {\dot Y}X-{\dot X}Y$ and ${\cal R}^2 =  X^2+Y^2$. 

Integrating the weak field equations for the metric outside of a finite spinning mass $M$, of angular momentum $J$,  leads to the weak field limit of the exterior Kerr solution in 
the usual inertial frame spherical co-ordinates ($t,r, \theta, \phi$) \cite{LnL} 
\begin{equation}
  g_{\mu\nu}=
  \left[ \begin{array}{cccc}
   1-r_s/r & 0 & 0 & ar_s/r\\
   0 & -1/(1-r_s/r) & 0 & 0 \\
   0 &  0  &  -r^2 & 0 \\    
   ar_s/r & 0 & 0 &-r^2\sin^2\theta \\
  \end{array}  \right]
\label{metric}
\end{equation} 
where $r_s = 2GM/c^2$ and $a=J/(Mc)$ as before. Next we consider orbits of a much smaller mass ($m<<M$) in this spacetime. By symmetry, a planar, circular orbit has a constant 4-velocity $\frac{{\rm d}x^\alpha}{{\rm d}s} = u^\alpha = (u^0, 0, 0, u^3)$ that solves the equations of motion. The components can be recast into two constants, akin to the energy, ${\tilde E}$, and the angular momentum, ${\tilde L}$, as ascribed to the mass $m$ by an asymptotic observer, 
\begin{equation} 
u^0 = \frac{r{\tilde E}}{r-r_s}-\frac{ar_s{\tilde L}}{r^2(r-r_s)} \qquad u^3 = \frac{ar_s{\tilde E}}{r^2(r-r_s)} + \frac{\tilde L}{r^2}
\label{u0u3}
\end{equation}

Therefore, in the sign convention of Eq.~(\ref{u0u3}), the expression for the angular velocity with respect to asymptotic time for a single center system is
\begin{equation} {\frac{{\rm d}\phi'}{{\rm d}t}} = {\dot \phi'} = {\frac{{\tilde L}/r^2+ar_s{\tilde E}/(r^2(r-r_s))}{r{\tilde E}/(r-r_s)-a{\tilde L}r_s/(r^2(r-r_s))}}
\label{phiPrime1}
\end{equation}
where $\phi'= \phi+\Omega$ is the angular co-ordinate in the lab frame (the inertial frame of Eq.~(\ref{u0u3}), as seen from the BH 'center') and $\Omega$ is the rotating frame rate. The Cartesian co-ordinates in the rotating frame are thus given by $X = r\cos(\phi)+X_1$ and $Y = r\sin(\phi)$. For purpose of clarity, it may be beneficial for the reader to think of the 'velocity', ${\dot \phi}$, as small, since for most of the orbits we expect ${\dot \phi'}\sim \Omega$. For this reason, in terms of the rotating frame angle $\phi' = \phi+\Omega t$ 
\begin{equation}
{\dot \phi} = {\frac{(X-X_1){\dot Y}-Y{\dot X}}{r^2}}. 
\label{phiPrime_dot}
\end{equation}
where again $X$ and $Y$ are Cartesian co-ordinates as seen in the frame rotating about the CM at angular frequency $\Omega$. Note that $r^2 = (X-X_1)^2 + Y^2$ is the (parameter) distance to the black hole.

We now expand Eq.~(\ref{phiPrime1}) out to leading order in $a$ (dropping ${\tilde\ }$'s for clarity),
\begin{equation}
{\dot \phi} = (L/E) {\frac{r-r_s}{r^3}} +a({\frac{r_s}{r^3}} + (L/E)^2{\frac{r_s(r-r_s)}{r^6}}) -\Omega
\label{phiDot}
\end{equation}

Since \cite{HW} have already developed the relativistic corrections to leading order without rotation, we only need to focus here on the new terms that arise when rotation is included. Thus, we develop the radial equation of motion to leading order in $a$,
\begin{equation}
    {\ddot r } -r ({\dot \phi'})^2  = -{\frac{r_s}{2r^2 }} + {\frac{ar_s}{r^2}}{\dot \phi'}
\label{radial1}
\end{equation}
As expected, rewriting Eq.~(\ref{radial1}) and Eq.~(\ref{phiDot}) in the $X$ and $Y$ co-ordinates leads to the classical limit of Eqs. (13) and (14) of Ref. \cite{HW}, amended only by terms proportional to $a$. For first and second derivatives, we have
\begin{equation}
r {\dot r} = (X-X_1){\dot X} + Y{\dot Y} \qquad  {\dot \phi'} -\Omega = {\frac{(X-X_1){\dot Y}-Y{\dot X}}{r^2}}
\label{dot}
\end{equation}
and also
\begin{eqnarray}
r^3 {\ddot r} &&= r^2 ((X-X_1){\ddot X} + Y {\ddot Y}) + Y^2 {\dot X}^2+ (X-X_1)^2{\dot Y}^2 \nonumber\\
&&-2(X-X_1)Y {\dot X} {\dot Y}
\end{eqnarray}
\begin{equation}
{\frac{\rm d}{{\rm d}t}} (r^2(\dot \phi' - \Omega)) = (X-X_1){\ddot Y}- Y{\ddot X}
\label{ddot}
\end{equation}
To further simplify this result, it is useful to define the function that arises in Eq.~(\ref{phiDot}); specifically, let $f(r) \equiv {\frac{r_s}{r^3}} +{\frac{(L/E)^2 r_s(r-r_s)}{r^6}}$ and denote $f' = \frac{{\rm d}f}{{\rm d}r}$. With this nomenclature, the time derivative of Eq.(\ref{phiDot}) becomes,  
\begin{eqnarray}
&&(X-X_1){\ddot Y} -Y{\ddot X} + 2\Omega ((X-X_1){\dot X}+Y{\dot Y})
 \nonumber \\
&& = (L/E)r_s {\dot r}/r^2 + a{\dot r} (2rf+r^2f') + \ldots
\label{A12}
\end{eqnarray}
where we are assuming that the secular evolution of the ratio $(L/E)$ in the full (restricted three body) problem leads to sub-dominant terms, since $m/M$ is assumed small. 
Furthermore, by using Eq.~(\ref{ddot}), the radial equation Eq.~(\ref{radial1}), in these co-ordinates becomes
\begin{eqnarray}
&&(X-X_1){\ddot X}+Y{\ddot Y} - r^2\Omega^2 - 2\Omega ((X-X_1){\dot Y}-Y{\dot X}) 
 \nonumber \\
&&= -{\frac{r_s}{2r}} + {\frac{ar_s}{r}}(\Omega + ((X-X_1){\dot Y}-Y{\dot X})/r^2)+\ldots
\label{A13}
\end{eqnarray}
(where again the ``$\ldots$'' represent all the higher order relativistic terms independent of $a$.)
Next, by forming linear combinations of Eq.~(\ref{A12}) and Eq.~(\ref{A13}), the equations of motion are combined into a form closer to that of Eqs. (13) and (14) of Ref. \cite{HW}. Additionally, transforming into the rotating frame centered about $(X,Y) =  (0,0)$, one has 
\begin{eqnarray}
{\ddot X} &&= 2\Omega{\dot Y}+\Omega^2 X -{\frac{{r_s}(X-X_1)}{2r^3}}
-{\frac{a}{r^2}} \biggl(Y {\dot r}(2rf+r^2f')\nonumber\\
&&-{\frac{(X-X_1) r_s}{r}}(\Omega +((X-X_1){\dot Y}-Y{\dot X})/r^2)  \biggr)+\ldots
\label{A14}
\end{eqnarray}
with the other linear combination being 
\begin{eqnarray}
{\ddot Y} &&=-2\Omega{\dot X}+\Omega^2 Y
-{\frac{{r_s}Y}{2r^3}}+{\frac{a}{r^2}} \biggl((X-X_1) {\dot r}(2rf+r^2f')\nonumber \\
 &&+{\frac{Y r_s}{r}}(\Omega +((X-X_1){\dot Y}-Y{\dot X})/r^2)  \biggr)+\ldots
\label{A15}
\end{eqnarray}

Eqs.~(\ref{A14}, \ref{A15}) now contain the leading '$a$'-dependent terms other than those from the change in $\Omega$ due to the pair's motion (see Eq.~(\ref{orbitalRate}) '$a$' dependence). The '$\ldots$' indicate the leading order relativistic terms that are already in $P$ and $Q$ of Eqs.~(\ref{prel}) and (\ref{qrel}) respectively. 

Regarding the function $f$, as defined in the text above Eq. (\ref{A12}), we note 
\begin{eqnarray}
{\dot r}(2rf+r^2f')&&= ((X-X_1){\dot X}+Y{\dot Y}) \nonumber \\ 
&&\times \bigl(-{\frac{r_s}{r^3}}-(L/E)^2 {\frac{r_s(3r-4r_s)}{r^6}}\bigr)
\label{A16}
\end{eqnarray}
Thus during the integration of Eqs.~(\ref{A14}) and (\ref{A15}), in which the LHS of Eq.~(\ref{A16}) appears, one must continually update the (in the full system) quasi-constant $L/E$  using Eq.~(\ref{phiDot}). Since the terms in Eqs.~(\ref{A14}, \ref{A15}) involving $f$ are already of order $a$, for  Eq.~(\ref{A16}) it suffices to use, 
\begin{equation}
 (L/E) = {\frac{r^3}{(r-r_s)}} ({\dot \phi} + \Omega) = {\frac{r^3}{(r-r_s)}} ( {\frac{X{\dot Y}-Y{\dot X}}{r^2}} + \Omega) 
\label{A17}
\end{equation}
which we square and include in Eq.~(\ref{A16}) before finally using its LHS in Eqs.(\ref{A14}, \ref{A15}). 

In addition to integrating the equations of motion in time, it is straightforward to  numerically determine the linear stability of the system about the equilibrium positions in the rotating frame. To motivate this approach, we start with the familiar classical problem;  a particle subject to a potential ${\tilde V}({\vec x})$, for which we are looking for solutions that are 
static in some rotating frame, {\it i.e.} stable circular orbits. The generic two-dimensional lagrangian in inertial co-ordinates, ${\tilde x}, {\tilde y}$, is given by  ${\cal L} = \frac{1}{2} ({\dot {{\tilde x}}}^2 + {\dot {{\tilde y}}}^2) -{\tilde V}({\tilde x},{\tilde y})$ which, in the rotating frame,  becomes ${\cal L} = \frac{1}{2} ({\dot {x}}^2 + {\dot {y}}^2) + \Omega({\dot x}y - {\dot y}x) -{\tilde V}(x,y)+\frac{\Omega^2}{2}(x^2+y^2)$. 
Thus, in terms of the new potential $V = {\tilde V}-\frac{\Omega^2}{2} r^2$, the EOM become 
\begin{equation} 
{\ddot {\vec x}} -2{\vec \Omega}\times{\dot {\vec x}} + \nabla V = 0
\label{eom_rotframe}
\end{equation}
The full problem of concern here has the additional complexity that the relativistic terms in the
potential $V({\vec x},{\dot {\vec x}})$ are velocity dependent. For studying the stability of lagrange points in the rotating frame it  suffices to expand the EOM about the points (${\vec x^*}$) for which  $\nabla V|_{x^*}({\vec x}, 0) = 0$.  The velocities with respect to the rotating frame are zero at ${\vec x^*}$ to this order. Expanding to leading order as a linear system, we have
\begin{equation} 
{\ddot {\vec x}}  + {\bf {\cal S}}{\dot {\vec x}}+ {\bf {\cal H}}{\vec x} = 0
\label{eom_rotframe2}
\end{equation}
where ${\bf {\cal H}}_{ij} = \partial_i\partial_j V$ is the Hessian of the potential in the rotating frame, evaluated at the point ${\vec x^*}$ and ${\bf {\cal S}}_{ij} = 2\Omega\epsilon_{ij}+\frac{\partial \partial_j V}{\partial {\dot x}_i }$. 

Taking the ansatz $\vec x = {\vec A}e^{-\gamma t}$ for some non-zero constant vector ${\vec A}$  into Eq.~(\ref{eom_rotframe2}) then leads to a quartic equation for $\gamma$, 
\begin{eqnarray} 
\gamma^4 - ({\bf {\cal S}}_{xx}+&&{\bf {\cal S}}_{yy})\gamma^3 
+ ({\bf {\cal S}}_{xx}{\bf {\cal S}}_{yy}-{\bf {\cal S}}_{xy}{\bf {\cal S}}_{yx}+{\bf {\cal H}}_{xx} + {\bf {\cal H}}_{yy})\gamma^2 \nonumber \\
&&- ({\bf {\cal H}}_{xx}{\bf {\cal S}}_{yy}+{\bf {\cal H}}_{yy}{\bf {\cal S}}_{xx}-{\bf {\cal H}}_{xy}{\bf {\cal S}}_{yx}-{\bf {\cal H}}_{yx}{\bf {\cal S}}_{xy})\gamma \nonumber \\
&&+{\bf {\cal H}}_{xx}{\bf {\cal H}}_{yy} - {\bf {\cal H}}_{xy}{\bf {\cal H}}_{yx} = 0
\label{eigs_ahoy!}
\end{eqnarray} 
Lastly, to find the stability boundary, one need to only plot the (appropriate) roots of the discriminant Eq.~(\ref{eigs_ahoy!}). Next we describe the numerical evaluation of the full EOM Eqs.~(\ref{A14}) and (\ref{A15}) used to create the zero locus of the discriminant of Eq.~(\ref{eigs_ahoy!}) and also used to to compute the orbits in this case.

\section{\label{sec:3} Numerical Solution} 

\begin{figure}[t]
\includegraphics[width=86mm,scale=1.5]{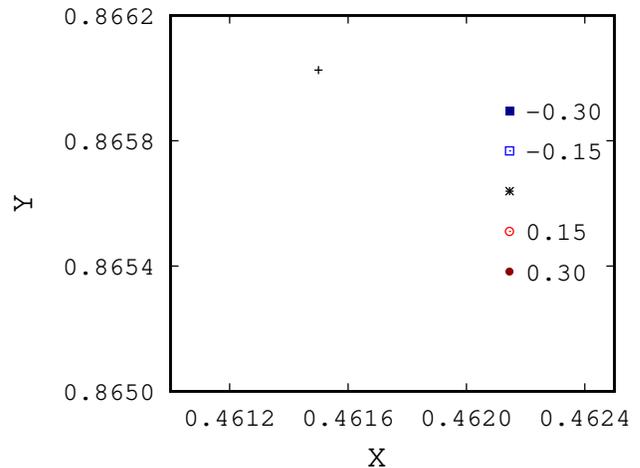}
\caption{The equilibrium points shift due to the relativistic correction terms. For $\mu = 0.0385$ the cross indicates the classical ($1/c^2 = 0$) location of L4. The "*" is the lagrange point for $c$ = 30 but with $a=0$. The other points are with this same $c$ value, but at the $a/a_{max}$ values indicated.} 
\label{lpoints} 
\end{figure}

In general, the classical triangular equilibrium point locations shift due to the relativistic correction terms. Therefore, by rewriting Eqs. (\ref{A14}) and (\ref{A15}) into a system of four, first order, nonlinear differential equations, the resulting vector field's zeroes were determined using GNU Scientific Library's (GSL's) multidimensional root-finding routine. The particular routine implemented a modified version of Powell’s Hybrid algorithm, but replaced calls to the Jacobian with finite difference approximations. As expected, $L_4$ and $L_5$ shift together, maintaining mirror symmetry across the line connecting the principal masses. Furthermore in the relativistic regime, the equilibrium points move nearly parallel to the principal axis toward the secondary mass. As prograde rotation increased, both $L4$ and $L5$ moved toward one another in the direction of the principal axis; however, retrograde orbital motion shifted the location of the points outwards (Fig.~\ref{lpoints}).

\begin{figure}[t]
\includegraphics[width=86mm,scale=1.5]{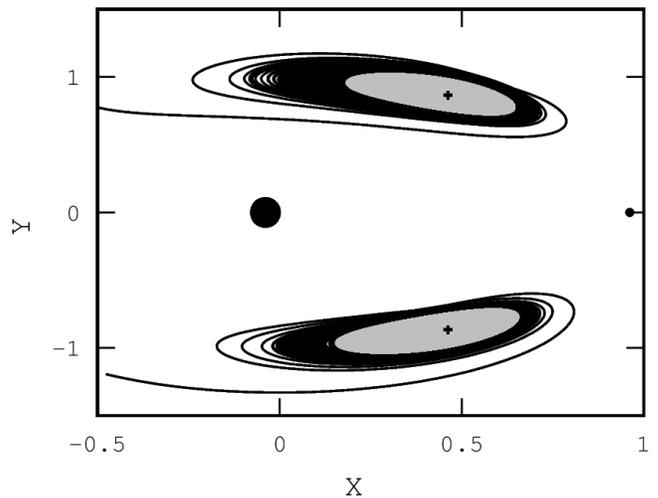}
\caption{\label{orbits}Typical results of numerically integrating the EOM with $a=0$. The center of mass of the system is located at the origin. The host mass is located at ($-\mu_2$,0) and is much heavier than the secondary mass at ($\mu_1$,0). The dark traces are representative of unstable orbits ($\mu>\mu^*_{classical}$) starting near $L4$ and $L5$. The light trace is similar, except for stable orbits ($\mu<\mu^*_{classical}$).} 
\end{figure}

The second portion of the code solved the orbital equations (Eqs. (\ref{A14}) and (\ref{A15})) for the motion of the test mass (Fig.~\ref{orbits}). The host and secondary mass separation remained constant at one unit, while the total mass of the system was also fixed at one in gravitational units ($G_N=1)$. Starting at locations near the equilibrium points, a test particle's trajectory was evolved using an explicit Runge-Kutta Prince-Dormand (8, 9) integrator for 100,000 orbits. The errors produced in each step were held within an absolute error bound of 1e-15.

\begin{figure}[t]
\includegraphics[width=86mm,scale=1.5]{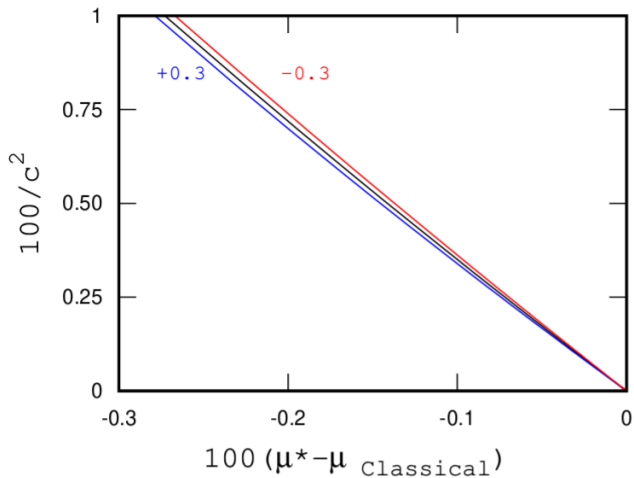}
\caption{ Computed critical $\mu$ boundary versus $1/c^2 = r_s/(2r)$ for different $J/M$ values for the host black hole. Each line corresponds to a particular critical binary; points above the lines are binaries that do not support stable $L4$, $L5$ orbits whereas those corresponding to points below the line do have stable $L4, L5$ orbits. The x-axis is the classical (Newtonian) limit. The $J/M = 0$ computed curve (middle) is a line with slope $-17\sqrt{69}/486 = -0.29056$ as indicated in the literature (\cite{D&P}, Eq.(11)). The surrounding lines, at the $a/a_{max}$ values indicated, show that retrograde (negative) $J/M$ binaries have stable $L4$ and $L5$ at larger mass ratios.} 
\label{mcrit_vsJ}
\end{figure}

The third part of the code numerically computes the second derivatives of the orbital differential equation's vector field, in both position and velocity, the ${\cal S}_{ij}$ and ${\cal H}_{ij}$ of Eq. (\ref{eom_rotframe2}), at the lagrange points $L_4$ and $L_5$ found by the first part of the code. These numerical derivatives are taken in a symmetric way to control numerical systematics. 


As a first check, both the orbital simulation as described above and the linear stability analysis (numerical evaluation of the discriminant of Eq.(\ref{eigs_ahoy!}) both reproduce the stability boundary for $L_4$ and $L_5$, namely, 
$\mu>\mu^*_{classical} =  \frac{1}{2}-\frac{\sqrt{69}}{18} \sim$  0.038521 in the 
classical limit $1/c^2 \rightarrow 0$,  the large distance limit of the full relativistic problem. A second check is provided by comparing the $a=0$ limit at nonzero $1/c^2$ from our code to the known analytical solution (Ref.\cite{D&P}). That comparison indicates not only that the stability boundary at $a=0$ should be a line, but that the slope of the line should be $ -17\sqrt{69}/486 \sim$ -0.29056, which in the numerical method described here reproduced to the leading 6 decimal places. 

Using the method we've developed to study stability near the classical limit but at finite $a=J/(Mc)$, we summarize the entire leading order relativistic stability boundary as (Fig~\ref{mcrit_vsJ}) 
\begin{equation} 
\mu^*(a,c) = \mu^*_{classical}-\frac{17\sqrt{69}}{972}\frac{r_s}{r}-0.0355 \frac{a}{r} + \ldots 
\label{stabilityEnvelope}
\end{equation}
Note that although we have derived this in the small $a$ limit, since $a<r_s/2$, we expect the rotation contribution to the stability boundary is always smaller in magnitude than the leading relativistic term. 

\section{\label{sec:4} Discussion} 
The slip of the positions of $L_4$ and $L_5$ with respect to the nominal classical locations has a consistent trend that correlates with stability\footnote{We are indebted to J. Schnittman for this observation and explanation}. The energy level sets in the rotating frame are tear-drop shaped around each lagrange point, with the steepest part of that potential at the section nearest the smaller of the two principal masses. Orbits near the lagrange points in the stable regime explore the tear-drop shaped region, and as one increases the mass ratio to the stability boundary, the gradient of the potential in the part of the tear-shaped region nearest the secondary increases. The association of instability with the orbits navigating a region with larger gradients holds in both the post-newtonian limit \cite{greeks,HW,D&P} and in the various dissipative variants of the classical problem \cite{Schnittman, Murray, SinghBello}. In the post-newtonian limit (with no rotation of the host mass) the shift of the lagrange points towards the secondary is, in fact, a type of relativistic kinematical focusing. 

Including the effects of rotation of the "host" mass $M$ into the problem, we find, due to frame dragging, information that a prograde orbit tends to shift $L_4$ and $L_5$ towards the axis between the bodies. Again, this increases the gradients of the potential near $L_4$ and $L_5$, thus making the system less stable. In a retrograde binary the opposite happens, leading to a greater stability near $L_4$, $L_5$.

The result that the lagrange points are less stable in a prograde system than in a retrograde system appear counterintuitive with respect to the single body case. Recall that for an isolated test mass revolving around a spinning black hole, for a given fixed parameter distance $r$, the prograde orbits have both higher frequency and smaller ISCO (innermost stable circular orbits)\cite{ISCO,ISCOspinning} than retrograde orbits. Recall further, the ISCO is the boundary between stable and unstable circular orbits for the relativistic case. Quite separately, in a generic parametric oscillator it is not unusual for (in the linearized picture) roots of the characteristic equation to merge with one another at or near zero frequency before the system becomes unstable (i.e. admit solutions of decaying amplitude). Both ISCO size and this eigenvalue flow thus indicate that the prograde system should be more stable than the retrograde one. 

We can relieve the tension between this qualitative picture of stability and that or our findings for $L_4$ and $L_5$ for a rotating host in a binary pair by comparing the system's bound state energy in the $M>>m$ limit. 
Note that  the energy of the orbital system depends on the rotation parameter $a$. A brief calculation indicates that the asymptotically accorded total system energy for a circular orbit binary in the limit that the black hole mass $M$ is much larger than the secondary (mass $m$) and in the limit of small $a$, is 
\begin{equation} 
E = M+\frac{m(1-2M/R-2Ma\omega/R)}{\sqrt{1-3M/R-6aM\omega/R}}
\label{EE}
\end{equation} 
where $\omega$ is the revolution rate which in this $M>>m$ limit in asymptotically inertial co-ordinates ($R\rightarrow \infty$), 
\begin{equation} 
\omega = \omega_0 \pm |a|\omega_0^2
\label{ww}
\end{equation} 
where $\omega_0^2 = G(M+m)/R^3$ and the $\pm$ is for prograde/retrograde orbital motion respectively. Combining Eqs.~(\ref{EE}) and (\ref{ww}) we learn that the total energy of the retrograde system is always smaller than that of the prograde system in this limit (large $R$), all other factors the same. We conjecture that this difference persists to all $R$. If so, the ISCO ordering (where by the test mass is captured by the black hole) and the finding here (whereby the test mass at $L_4$ or $L_5$ is generically thrown out of the system) are consistent, since it indicates that the retrograde system is more strongly bound than the equivalent prograde one. 

Astrophysically this suggests that  a rapidly spinning black hole host with a retrograde $m$ could only support stable $L_4$ and $L_5$ system for a particular annulus of circular orbits. Rather than always reducing the critical mass ratio for stability (as cited in earlier literature in the case of no rotation), relativistic effects due to the rotation of the host mass can actually increase the critical mass ratio (Fig~\ref{mcrit_vsJ}). If prograde orbital rotation is most likely astrophysically, the forgoing suggests that the critical ratio for Lagrange point stability happens for even smaller mass ratios than allowed classically. As known from earlier work, for a light secondary paired with a rapidly spinning black hole host, motions of masses at $L_4$ and $L_5$ stable at large distances will become unstable as the binary shrinks, and lead to ejection from the system for any rotation of the host mass.

%


\begin{acknowledgments}
We wish to thank J. Schnittman for sharing his insights into the dynamics of the non-rotating case in leading non-relativistic order and C. Johnson for discussion on the qualitative link between the findings and spacetime entropy change. 
\end{acknowledgments}


\bibliography{KerrLagrange}
\bibliographystyle{aipauth4-1}
\bibliographystyle{unsrtnat}
\include{output.bbl}

\end{document}